\documentclass[journal]{IEEEtran}

\usepackage{boxedminipage}
\usepackage{amsthm}
\usepackage{amssymb,amsmath,bm}
\usepackage{multirow}
\usepackage{multicol}
\usepackage{macros}
\usepackage{times}
\usepackage{graphicx}
\usepackage{epsfig}
\usepackage{color}
\usepackage{array}
\usepackage{tabularx}
\usepackage{subfig}

\ifCLASSINFOpdf
\else
\fi

\begin{document}
\title{Reducing Reconciliation Communication Cost \\ with Compressed Sensing}

\author{H. T. Kung and Chia-Mu~Yu}

\markboth{Journal of \LaTeX\ Class Files,~Vol.~6, No.~1, January~2010}%
{Shell \MakeLowercase{\textit{et al.}}: Bare Demo of IEEEtran.cls for Journals}

\maketitle

\begin{abstract}
We consider a reconciliation problem, where two hosts wish to synchronize their respective sets. Efficient solutions for minimizing the communication cost between the two hosts have been previously proposed in the literature. However, they rely on prior knowledge about the size of the set differences between the two sets to be reconciled. In this paper, we propose a method which can achieve comparable efficiency without assuming this prior knowledge. Our method uses compressive sensing techniques which can leverage the expected sparsity in set differences. We study the performance of the method via theoretical analysis and numerical simulations.

\end{abstract}


\IEEEpeerreviewmaketitle
\section{Introduction}\label{sec: Introduction}
Set reconciliation occurs naturally. For example, routers may need to reconcile their routing tables and files on mobile devices may need to be synchronized with those in the cloud. The reconciliation problem is to find the set differences between two distributed sets. Here, the set difference for a host is defined as the set of elements that the host has but the other host does not. Once two hosts can find their respective set differences, each can use the information to solve the reconciliation problem by adding its difference set to the other or removing it from its own set to reconcile the two sets to their union or intersection, respectively. In this paper, for presentation simplicity, we consider a simpler case that a host just reconcile its set to the same as the set that the other host currently possesses.

We describe the problem we wish to solve in mathematical notation. Suppose that there are two hosts, $A$ and $B$, which possess two sets, $S_A$ and $S_B$, respectively. The elements of $S_A$ and $S_B$ are from a set $U\subseteq \mathbb{N}$. The difference sets for $A$ and $B$ are $\Delta_A=S_A\setminus S_B$ and $\Delta_B=S_B\setminus S_A$, respectively. For example, if $A$ has $S_A=\{1,2,3\}$ and B has $S_B=\{2,3,4\}$, then we have $\Delta_A=\{1\}$ and $\Delta_B=\{4\}$. We denote the size of a set $S$ by $|S|$. To ease the presentation, we assume throughout the paper that $|S_A|$, $|S_B|\leq n$ and $d=|\Delta_A|+|\Delta_B|\leq n$ for some positive integer $n$. The method proposed in this paper can be naturally extended to the case of $n<d\leq 2n$ by simply increasing the space allocation from $2n$ to $4n$ (described in Sec. \ref{sec: CS-IBLT}).

In the reconciliation problem, the two hosts wish to reconcile their sets, by making them identical. For example, $B$ can update $S_B$ by adding elements in $\Delta_A$ to $S_B$ and removing elements in $\Delta_B$ from $S_B$. This means, in the above example, once $B$ knows $\Delta_A=\{1\}$ and $\Delta_B=\{4\}$, $B$ performs the operation of $(S_B\cup \Delta_A)\setminus \Delta_B$. Consequently, the reconciliation is accomplished.

In solving the reconciliation problem, we are mainly concerned with the communication cost, the number of elements required to be transmitted between the two hosts.

\subsection{Related Work}\label{sec: Related Work}
A straightforward method of solving the reconciliation problem is that host $A$ sends his entire set $S_A$ to host $B$. After that, $B$ can check and identify the set differences between $S_A$ and $S_B$. Obviously, the communication cost for this method is $|S_A|$.

A more efficient but probabilistic method is to utilize Bloom filter \cite{b70}. More specifically, host $A$ constructs a Bloom filter by inserting the elements in $S_A$ to the Bloom filter and then sending the Bloom filter to $B$. With the received Bloom filter, $B$ can check if the elements in $S_B$ is in the filter and thus can identify $\Delta_B$ with some probability that not all these elements are identified due to hash table collisions in the Bloom filter. Similar queries made for the remaining elements in $U$ can be used to identify $\Delta_A$ with some probability that extra elements are identified due to hash table collisions in the Bloom filter. To lower false identifications, the size of Bloom filter needs to be proportional to $n$. Therefore, the communication cost of this Bloom filter approach is still asymptotically the same as the straightforward method.

Minsky \emph{et al}. \cite{mtz03} developed a characteristic polynomial method. In this method, $A$ sends several evaluated values of the characteristic polynomial $c_{S_A}$ to $B$, where $c_{S_A}$ is defined as $c_{S_A}=\prod_{i=1}^{|S_A|}(Z-x_A^i)$ with $x_A^i$'s being elements in $S_A$. Host $B$ does similar evaluation based on its own characteristic polynomial $c_{S_B}$. By \emph{rational interpolation}, $B$ can derive $c_{S_A}$ and thus recover the set differences based on $c_{S_A}$'s and $c_{S_B}$'s evaluated values. Here, given $d_1+d_2+1$ pairs of $(k_i,f_i)$, rational interpolation is to find a $f=\frac{P}{Q}$ satisfying $f(k_i)=f_i$ for each pair $(k_i,f_i)$, where the polynomials $P$ and $Q$ are of degrees $d_1$ and $d_2$, respectively.

Observe that $\frac{c_{S_A}}{c_{S_B}}=\frac{c_{S_A\cap S_B}\cdot c_{\Delta_A}}{c_{S_A\cap S_B}\cdot c_{\Delta_B}}=\frac{c_{\Delta_A}}{c_{\Delta_B}}$. $A$ sends evaluated values of $c_{S_A}$ to $B$, and $B$ calculates the value of $\frac{c_{\Delta_A}}{c_{\Delta_B}}$ at each predetermined evaluation point. Once $\frac{c_{S_A}}{c_{S_B}}$ can be recovered from the evaluated values of $\frac{c_{\Delta_A}}{c_{\Delta_B}}$, the set differences can be obtained by finding the roots of $c_{\Delta_A}$ and $c_{\Delta_B}$. 

A concrete example in \cite{mtz03} shows how this characteristic polynomial method works. Suppose that $S_A=\{1,9,28,33,53,61\}$, $S_B=\{1,9,10,28,53\}$, the prior knowledge about $d$ is available, the evaluation points $\{0,1,2,3\}$ have been predetermined, and a proper finite field $\mathbb{F}_{97}$ has been chosen. Under such conditions, $c_{S_A}$ and $c_{S_B}$ can be formulated as $(Z-1)(Z-9)(Z-28)(Z-33)(Z-53)(Z-61)$ and $(Z-1)(Z-9)(Z-10)(Z-28)(Z-53)$, respectively. The evaluations of $c_{S_A}$ and $c_{S_B}$ at four evaluation points are $\{41,85,65,81\}$\footnote{A particular treatment needs to be taken on the evaluation point $1$, but we omit the detail in this paper.} and $\{9,14,51,46\}$ over $\mathbb{F}_{97}$, respectively. The values of $\frac{c_{S_A}}{c_{S_B}}$ are therefore $\{\frac{41}{9},\frac{85}{14},\frac{65}{51},\frac{81}{46}\}=\{80,13,26,84\}$. From rational interpolation's perspective, the value $d_1+d_2$ corresponds to the size $d$of set differences and $\{(k_i,f_i)\}$ corresponds to $\{(0,80),(1,13),(2,26),(3,84)\}$ of size $d_1+d_2+1=4$. The interpolated $f=\frac{Z^2-94Z+73}{Z-10}$, where the roots of numerator are $33$ and $61$ and the root of denominator is $10$, can be used to derive the set differences between $S_A$ and $S_B$. An issue in this reconciliation case is that only the size of set differences, instead of the individual $d_1$ and $d_2$, is known and so rational interpolation cannot be applied directly. Nevertheless, a formula is given in \cite{mtz03} to the estimates of $d_1$ and $d_2$ based only on the size of set differences. Despite its algebraic computation over finite fields, a notable feature of this method is that the communication cost is only dependent on $d$, instead of $n$, due to the use of interpolation.

Very recently, Goodrich and Mitzenmacher \cite{mm} developed a data structure, called invertible Bloom lookup table (IBLT), to address the reconciliation problem. IBLT can be thought of as a variant of counting Bloom filter \cite{fcab00} with the property that the elements inserted to Bloom filter can be extracted even under collision. With the use of IBLT, the reconciliation problem can be solved in approximately $2d$ communication cost under the assumption that $d$ is known in advance.

\subsection{Research Gap and Contribution}\label{sec: Research Gap and Contribution}
The aforementioned straightforward method and Bloom filter approach incur a large amount of communication cost when $S_A$ is of large size. On the other hand, characteristic polynomial method and IBLT are efficient only when prior knowledge about $d$ is available. Without this prior knowledge, the computation overhead of the characteristic polynomial method can be as large as $O(n^4)$. IBLT need to be repeatedly applied with progressively increasing $d$, incurring a wasted communication cost which can be as large as $O(n\log n)$.

We propose an algorithm, called CS-IBLT, which is a novel combination of compressed sensing (CS) and IBLT, enabling the reconciliation problem to be solved with $O(d)$ communication cost even without prior knowledge about $d$. A distinguished feature of CS-IBLT is that the number of transmitted messages changes with adapt to the value of $d$, instead of the conventional wisdom that the correct $d$ must be estimated first. Notably, this adaptive feature is attributed to the use of CS.

\section{Proposed Method}
First, we briefly review compressed sensing (CS) and invertible Bloom lookup table (IBLT) in Sec. \ref{sec: Compressed Sensing} and Sec. \ref{sec: Invertible Bloom Lookup Table}, respectively. Then, we describe our proposed CS-IBLT algorithm in Sec. \ref{sec: CS-IBLT}. We provide analysis and comparison between IBLT and CS-IBLT in Secs. \ref{sec: Analysis} and \ref{sec: Comparison}.

\subsection{Compressed Sensing}\label{sec: Compressed Sensing}
Suppose that $x$ is a $s$-sparse vector of length $n$ with $s\ll n$. That is, only $s$ nonzero components can be found in $x$. A standard compressed sensing (CS) formulation is $y=\Phi x$, where $y\in \mathbb{R}^m$ and $\Phi\in \mathbb{R}^{m\times n}$, with $m\ll n$, are called measurement vector and measurement matrix, respectively. CS states that if $\Phi$ is a random matrix satisfying the restricted isometry property and $m$ is greater than $cs\log\frac{n}{s}$ for some constant $c$ \cite{crt06}, then $x$ can be reconstructed based on $y$ with high probability. The vector $x$ can be reconstructed by $\ell_1$-minimization as follows: \begin{align} x^*=\underset{y=\Phi x}{\operatorname{argmin}}||x||_{\ell_1}.\end{align}

\subsection{Invertible Bloom Lookup Table}\label{sec: Invertible Bloom Lookup Table}
An invertible Bloom lookup table (IBLT) is composed of a $b\times 2$ array, $IBLT$, with $k$ hash functions, $h_1(\cdot)$, $\dots$, $h_k(\cdot)$. It supports three operations\footnote{As IBLT is designed originally for storing key-value pairs, it actually supports GET operation. The purpose of GET is to return the value for a given key. Since we do not deal with key-value pairs, we omit the description of the GET operation for the ease of presentation.}, INSERT, DELETE, and LIST-ENTRIES. Suppose that $e$ is a numeric value. To insert an element $e$ with the INSERT operation, $IBLT[h_i(e),1]$ is increased by $e$ and $IBLT[h_i(e),2]$ is increased by $1$, for all $1\leq i\leq k$. The deletion of an element $e$ with the DELETE operation is operated by decreasing $IBLT[h_i(e),1]$ by $e$ and decreasing $IBLT[h_i(e),2]$ by $1$. The second column of IBLT can be treated as a counting Bloom filter \cite{fcab00}. LIST-ENTRIES is used to dump all elements currently stored in IBLT. It works by searching for the position $1\leq i\leq b$ where $IBLT[i,2]=1$. If such $i$ is found, the corresponding $IBLT[i,1]$ is listed and operation DELETE($IBLT[i,1]$) is performed. The above search-and-delete procedure is repeatedly performed until no such $i$ can be found. With this search-and-delete procedure, elements under collision can still be extracted. The LIST-ENTRIES operation fails if the resultant IBLT is not empty. It succeeds otherwise. Goodrich and Mitzenmacher show in \cite{mm} that to accommodate $n$ elements, the length $b$ of IBLT needs to be greater than $1.2n$ when $k$\footnote{When $k=4,5,6$, and $7$ are used, approximately $1.3n$, $1.4n$, $1.6n$, and $1.7n$ should be allocated, respectively. The rationale behind this is that for fixed IBLT size, larger $k$ implies more collision. To be able to perform the element extraction, collision cannot too much although collision is allowed in IBLT. Thus, when larger $k$ is used, more space allocation is required.} is selected to be $3$. This makes sure the LIST-ENTRIES fails with negligible probability.

\subsection{CS-IBLT}\label{sec: CS-IBLT}
Recall that $S_A$ and $S_B$ are two sets of length $n$. Under CS-IBLT, host $A$ first constructs an IBLT, $IBLT_A$, of length $2n$ by inserting each element in $S_A$ to $IBLT_A$. (The choice of $2n$ will be described in Sec. \ref{sec: Analysis}.) Host $A$ then constructs a random measurement matrix $\Phi$ of dimension $m\times 2n$ satisfying the restricted isometry property mentioned in Sec. \ref{sec: Compressed Sensing}. $A$ calculates $y_A=\Phi\cdot IBLT_A$. $y_A$ is thus an array of dimension $m\times 2$. Afterwards, $A$ repeatedly sends the rows of $y_A$ to $B$ continuously until it receives a positive acknowledgement from $B$ (described below).

Host $B$ constructs $IBLT_B$, $\Phi$, and $y_B$ in a similar manner. Note that with a seed commonly shared between $A$ and $B$, their generated $\Phi$ can be the same for each row. Denote the $i$-th row of $y_A$ by $y_A^i$. Once receiving the $i$-th row $y_A^i$ of $y_A$, $B$ performs CS recovery on $[y_A^1-y_B^1\mbox{ }y_A^2-y_B^2\mbox{ } \cdots\mbox{ } y_A^i-y_B^i]^T$. By CS recovery on $[y_A^1-y_B^1\mbox{ }y_A^2-y_B^2\mbox{ } \cdots\mbox{ } y_A^i-y_B^i]^T$, we mean that $\ell_1$-minimization is applied to the two columns in $[y_A^1-y_B^1\mbox{ }y_A^2-y_B^2\mbox{ } \cdots\mbox{ } y_A^i-y_B^i]^T$ separatively. Because the entries in $IBLT_A$ and $IBLT_B$ are assumed to be integers, quantization is applied to the recovered result. Suppose that $B$ obtains a recovery result $\widehat{IBLT}_{A-B}$ after $\ell_1$-minimization is applied to $[y_A^1-y_B^1\mbox{ }y_A^2-y_B^2\mbox{ } \cdots\mbox{ } y_A^i-y_B^i]^T$. $B$ then proceeds to the LIST-ENTRIES operation on $\widehat{IBLT}_{A-B}$ and checks whether the LIST-ENTRIES operation succeeds or not. If the LIST-ENTRIES operation succeeds, $B$ sends a positive acknowledgment meaning "stop sending more measurements" to $A$, and host B reconciles $S_B$ with $S_A$, with the $\Delta_A$ and $\Delta_B$ extracted from $\widehat{IBLT}_{A-B}$. If the LIST-ENTRIES operation fails, $B$ waits for the next measurement $y_A^{i+1}$ and again performs the above operations on $y_A^1$ through $y_A^{i+1}$.

The above setting and procedures remain the same in the case of $n<d\leq 2n$ except that $IBLT_A$ and $IBLT_B$ of length at most $4n$ are needed instead. Note that $4n$ corresponds to the extreme case of $d=2n$.

\begin{figure}[hdt]
\centering
\includegraphics[width=9cm]{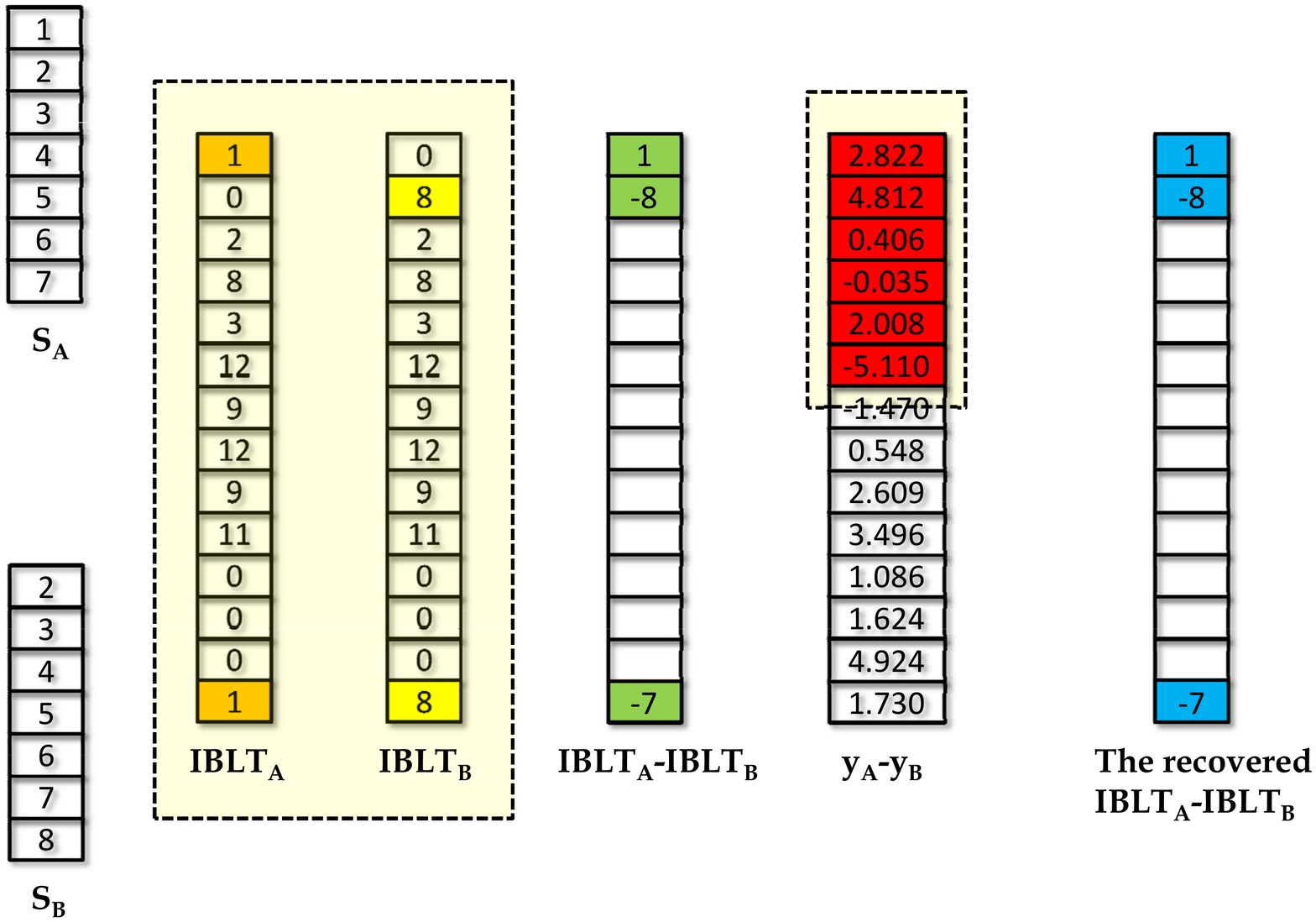}
\caption{An illustration of CS-IBLT.} \label{fig: CS-IBLT}
\end{figure}

Figure \ref{fig: CS-IBLT} illustrates how CS-IBLT works. Hosts $A$ and $B$ possess $S_A=\{1,2,\dots,7\}$ and $S_B=\{2,3,\dots,8\}$, respectively. In the following, we omit the second column of IBLT in our CS-IBLT algorithm for representation simplicity. That is, we omit the counting Bloom filter part. Observe  that $\Delta_A=\{1\}$, $\Delta_B=\{8\}$, and $d=2$. Note that because of $n=7$, IBLTs are of length $14$. This corresponds to the requirement in Sec. \ref{sec: CS-IBLT} that IBLTs of length $2n$ need to be allocated. Suppose that $k=2$ hash functions are used in the IBLT in CS-IBLT. $IBLT_A$ and $IBLT_B$ are derived according to the hash positions and then $IBLT_A-IBLT_B$ is calculated. With CS-IBLT, $A$ only needs to send the first $6$ entries in $y_A$ to $B$. That is, only six entries of $y_A-y_B$ are sufficient for $B$ to exactly recover the $IBLT_A-IBLT_B$. From the recovered $IBLT_A-IBLT_B$, $\widehat{IBLT_{A-B}}$, we can extract $1$ and $-8$ according to the IBLT principles in Sec. \ref{sec: Invertible Bloom Lookup Table}. Based on the rule described in Sec. \ref{sec: Analysis}, $B$ knows that $\Delta_A=\{1\}$, $\Delta_B=\{8\}$.

\subsection{Analysis}\label{sec: Analysis}
The following is the key relationship behind our proposed CS-IBLT algorithm is: \begin{align} y_A-y_B=\Phi(IBLT_A-IBLT_B).\end{align} The CS recovery based on $y_A-y_B$ can generate an approximation $\widehat{IBLT}_{A-B}$ of $IBLT_A-IBLT_B$. When the number $m$ of measurements is sufficient in the CS recovery, $\widehat{IBLT}_{A-B}$ is nearly identical to $IBLT_A-IBLT_B$. Based on the principles of IBLT construction, $IBLT_A-IBLT_B$ can be thought of as an IBLT with elements in $\Delta_A$ and in $\bar{\Delta}_B$, where $\bar{\Delta}_B$ is defined as the set $\{0-e|e\in \Delta_B\}$. Thus, $B$ first lists all the elements in $\widehat{IBLT}_{A-B}$. Those positive elements are categorized as $\Delta_A$ and those negative ones are categorized as $\bar{\Delta}_B$.

On the other hand, when the number $m$ of measurements is insufficient for the exact recovery of $IBLT_A-IBLT_B$. That is, $\widehat{IBLT}_{A-B}$ is significantly deviated from $IBLT_A-IBLT_B$, $B$ will be aware of this failed recovery because after the LIST-ENTRIES operation is applied to such $\widehat{IBLT}_{A-B}$, the LIST-ENTRIES operation fails with high probability. Note that the reconstructed array $\widehat{IBLT}_{A-B}$ behaves like a random one when an insufficient number of measurements is used. The LIST-ENTRIES operation is unlikely to be successful on a random array. Therefore, the decoding procedure will proceed with high probability until $\widehat{IBLT}_{A-B}\approx IBLT_A-IBLT_B$ is achieved.

The number of measurements required to recover $IBLT_A-IBLT_B$ determines the communication cost of CS-IBLT. Recall that we are interested in recovering $IBLT_A-IBLT_B$ from $y_A-y_B=\Phi(IBLT_A-IBLT_B)$, and the theory of CS states that the number of required measurements can be as small as $cs\log\frac{n}{s}$, where $s$ is the number of nonzero entries in the vector to be recovered. Observe that the IBLT, $IBLT_A-IBLT_B$, is constructed by adding elements in $S_A$ and removing elements in $S_B$. Based on the IBLT principles in Sec. \ref{sec: Invertible Bloom Lookup Table}, the elements commonly shared between $A$ and $B$, which are the elements in $(S_A\cup S_B)\setminus(\Delta_A\cup \Delta_B)$, will be eliminated and only the elements in the set difference $\Delta_A\cup \Delta_B$ remain in $IBLT_A-IBLT_B$. Recall that $cs\log \frac{n}{s}$ measurements are needed for accurate CS recovery, where $s$ is the number of nonzero elements. Thus, as the vector to be recovered is $\widehat{IBLT}_{A-B}$ with at most $kd$ nonzero entries, $\min\{2n,ckd\log\frac{n}{kd}\}$ measurements are sufficient for the CS recovery, where $k$ and $d$ denote the number of hash functions used in IBLT and the inherent size of set differences, respectively.

As reported in \cite{mm}, the length of IBLT with $n$ elements should be at least $1.2n$ to ensure the successful execution of the LIST-ENTRIES operation in the case of $k=3$. However, the value of $1.2n$ is estimated based on an inherent assumption that the inserted elements are all positive. Based on the IBLT principles in Sec. \ref{sec: Invertible Bloom Lookup Table}, $IBLT_A-IBLT_B$ can be regarded as an IBLT with elements of $\Delta_A$ and $\bar{\Delta}_B$. Since there could be some negative elements in $\Delta_A$ and $\bar{\Delta}_B$, we suggest to use $2n$, rather than $1.2n$, according to our empirical experience.

\subsection{Comparison}\label{sec: Comparison}
In the case that prior knowledge about $d$ is unavailable, the use of IBLT incurs a large amount of wasted communication. In particular, a reasonably first guess is $\hat{d}=\frac{n}{2}$, and host $A$ sends IBLT of size $2\hat{d}$ to $B$. If the real $d$ is smaller then $\hat{d}$, $B$ can obtain $\Delta_A$ and $\Delta_B$ successfully. Essentially, $2\cdot d$ communications are sufficient for finding the set differences and this means that we incur unnecessary communication cost which can be as large as $2\cdot \frac{n}{2}-2\cdot 1=n-2$. This extreme case occurs when $d=1$.

If the real $d$ is greater than $\hat{d}$, then the LIST-ENTRIES operation will be failed, and $B$ keeps waiting for the subsequent measurements from $A$. This time, $A$ adopts a binary search-like approach to progressively have next $\hat{d}=\frac{3}{4}n$. Afterwards, hosts $A$ and $B$ repeat the above procedures until $B$ can empty $\widehat{IBLT}_{A-B}$. In the extreme case of $d=n$, $2(\frac{n}{2}+\frac{3n}{4}+\dots)=O(n\log n)$ communication cost is required. This performance is even worse than that of straightforward method in which $S_A$ is sent to $B$ directly.

On the other hand, in the case of $d=1$, if CS-IBLT is used, since the array $IBLT_A-IBLT_B$ is very sparse (approximately only $d\cdot k=k$ nonzero entries), only a very small number of measurements are needed. In the case of $d=n$, $2n$ measurements are sufficient for the CS recovery in CS-IBLT. Such communication cost occurs when all of the rows of $y_A$ are transmitted.

\section{Numerical Experiments}\label{sec: Numerical Experiments}
\begin{figure}[ht!]
\centering
\subfloat[]{\label{fig: k2n200}\includegraphics[width=0.25\textwidth]{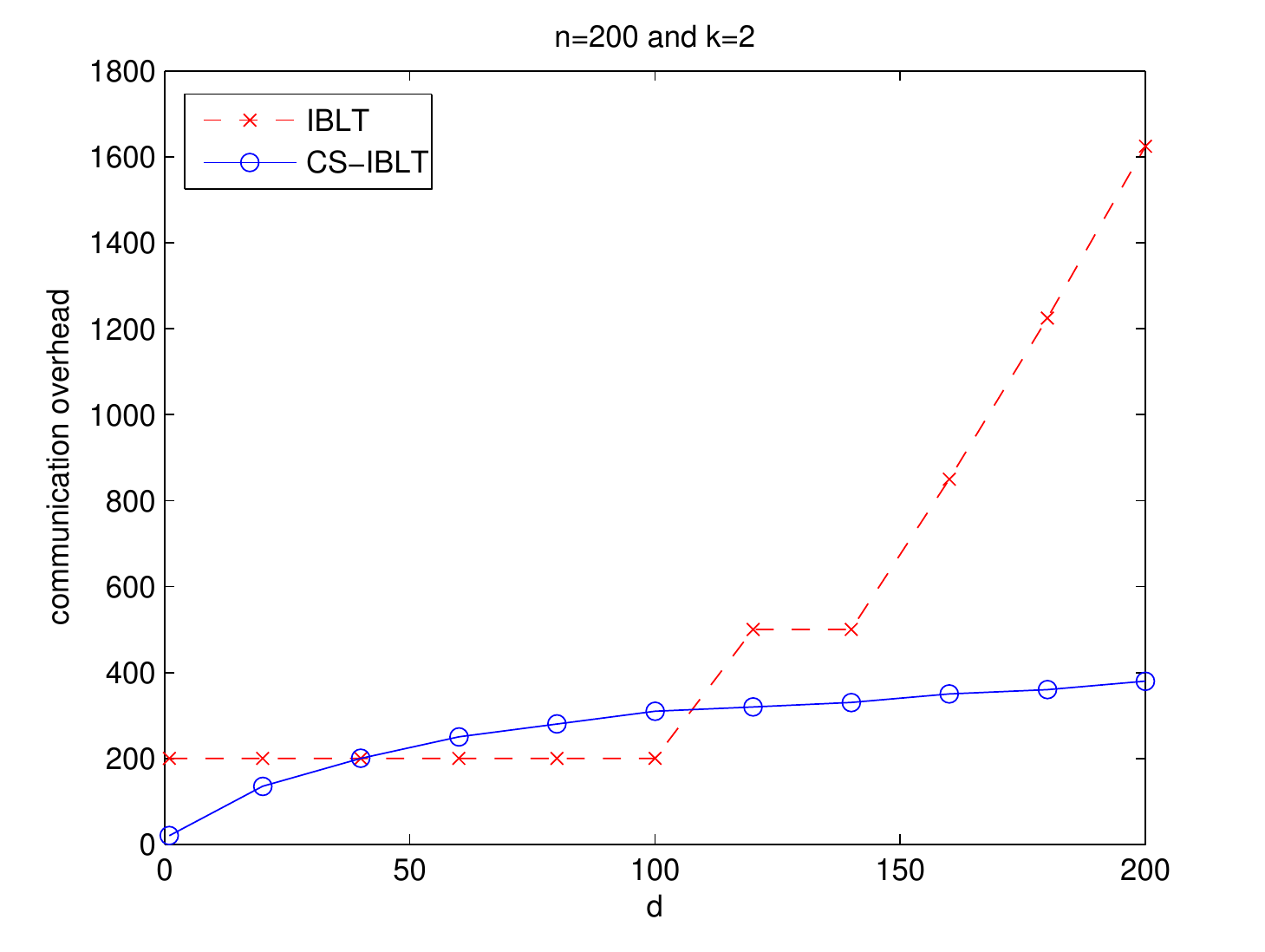}}
\subfloat[]{\label{fig: k2n1000}\includegraphics[width=0.25\textwidth]{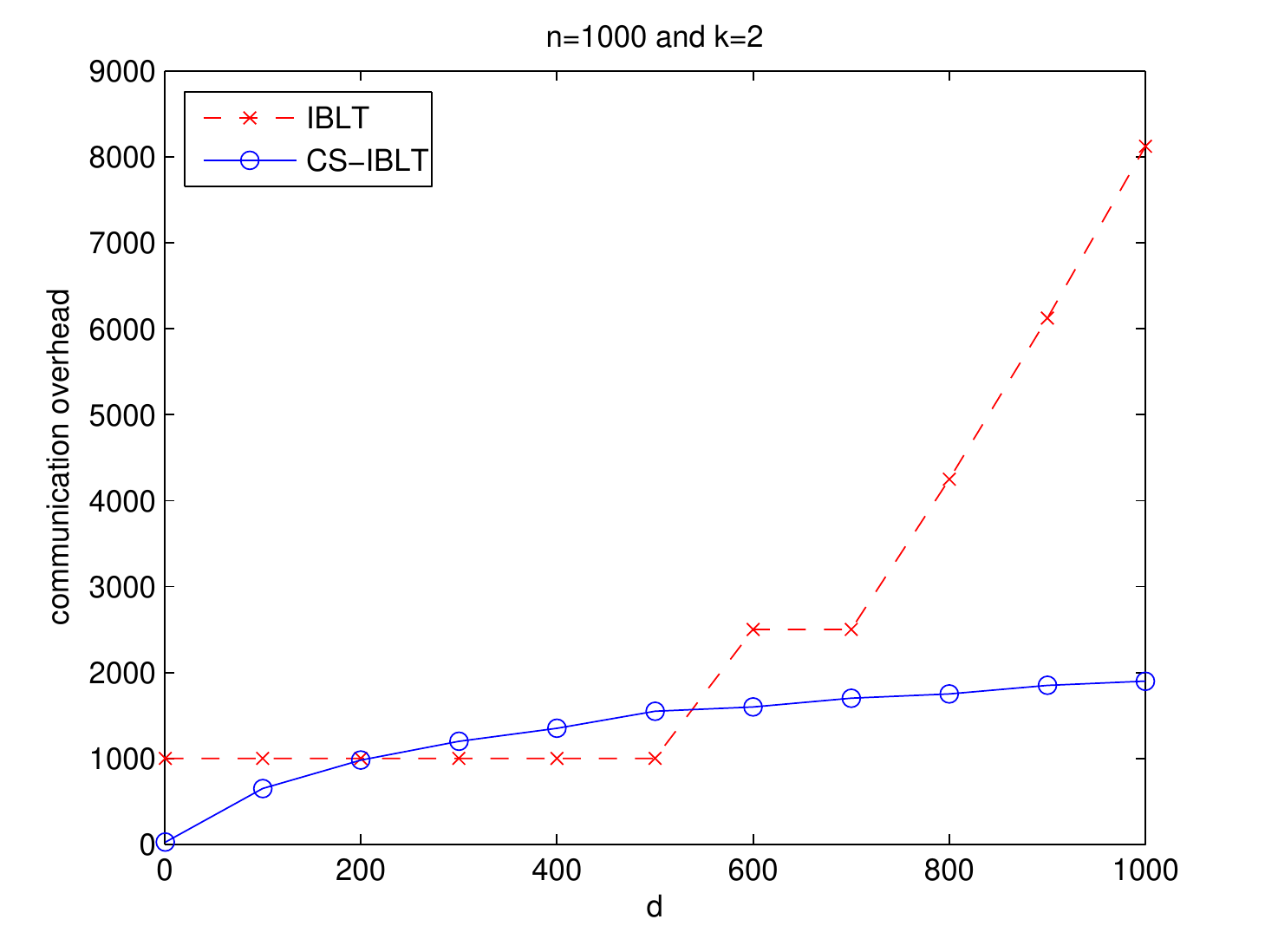}}
\caption{The size of set differences v.s. communication cost (a) $n=200$ and $k=2$ (b) $n=1000$ and $k=2$.} \label{fig: communication cost}
\end{figure}

In this section we demonstrate and compare the performance of IBLT and CS-IBLT via numerical experiments. Figure \ref{fig: communication cost} compares the performance of both methods under the assumption that prior knowledge about $d$ is not available.

In these experiments, $k=2$ hash functions are used in both IBLT and CS-IBLT. In CS-IBLT, the random measurement matrix $\Phi$ is Gaussian distributed. In Figure \ref{fig: k2n200}, $|S_A|=|S_B|=n=200$ and $d$ is varied from $1$ to $200$. One can see in Figure \ref{fig: k2n200} that communication cost of CS-IBLT increases as $d$ increases due to the fact that the larger $d$ implies more nonzero entries in $IBLT_A-IBLT_B$. In essence, the procedures in CS-IBLT here are roughly like applying CS measurement matrix to a $kd$-sparse array $IBLT_A-IBLT_B$ and then deriving the CS recovered array $\widehat{IBLT}_{A-B}$. On the other hand, in IBLT, because no prior knowledge about $d$ can be used, the guessed $d$, $\hat{d}=\frac{n}{2}$, is used initially. This choice of $\hat{d}$ enables $B$ to decode the received IBLT, resulting in a flat curve from $d=1$ to $d=100$. Similar observations can be made in Figure \ref{fig: k2n1000}.

CS-IBLT shows its main advantage when $d$ is relatively small and large. In the case of small $d$, the overestimated $\hat{d}$ incurs unnecessary communication but different measurements are adaptively transmitted one by one in CS-IBLT. The sending stops immediately after the successful recovery of $IBLT_A-IBLT_B$. In the case of large $d$, several underestimated $\hat{d}$ in IBLT incurs useless communication but because of its adaptive property, even in the worst case, $2n$ measurements can enable the successful recovery of $IBLT_A-IBLT_B$. CS-IBLT is inferior to IBLT only in the case of moderate $d$, which means that the initially guessed $d$, $\hat{d}$, is pretty close to the real $d$. The rationale behind this is that the communication cost of CS-IBLT is still limited by the theory of CS. That is, it is still dependent on $n$. However, if $\hat{d}\approx d$, we can think that IBLT with prior knowledge about $d$ is utilized, resulting in only $2d$ communication. Hence, in such cases, CS-IBLT is less efficient than IBLT in terms of communication cost.

\section{Conclusion}\label{sec: Conclusion}
We present a novel algorithm, CS-IBLT, to address the reconciliation problem. According to our theoretical analysis and numerical experiments, CS-IBLT is superior to the previous methods in terms of communication cost in most cases under the assumption that no prior information is available.

\mbox{ }\\

\noindent
{\bf Acknowledgment\/}:
Chia-Mu Yu was supported by NSC98-2917-I-002-116.

\end{document}